\documentclass[aps,prb,reprint,superscriptaddress]{revtex4-1}

\usepackage{graphicx}
\usepackage{color}

\begin{document}

\title{Electrical transport properties of bulk tetragonal CuMnAs}

\author{J. Voln\'y} 
\affiliation{Charles University,
Faculty of Mathematics and Physics, Department of
Condensed Matter Physics, Ke Karlovu 5, Praha 2, CZ--12116, Czech Republic}
\author{D. Wagenknecht} 
\affiliation{Charles University,
Faculty of Mathematics and Physics, Department of
Condensed Matter Physics, Ke Karlovu 5, Praha 2, CZ--12116, Czech Republic}
\author{J. \v Zelezn\'y}
\affiliation{Institute of Physics, Academy of Science of the Czech Rep.,
  Cukrovarnick\'a 10, Praha 6, CZ--16253}
\author{P. Harcuba}
\affiliation{\hbox{Charles
University, Faculty of Math. and Physics, Dep. of Material Physics,
Ke Karlovu 5, Praha 2, CZ--12116}}
\author{E.~Duverger--Nedellec}
\affiliation{Charles University,
Faculty of Mathematics and Physics, Department of
Condensed Matter Physics, Ke Karlovu 5, Praha 2, CZ--12116, Czech Republic}
\author{R.H.~Colman}
\affiliation{Charles University,
Faculty of Mathematics and Physics, Department of
Condensed Matter Physics, Ke Karlovu 5, Praha 2, CZ--12116, Czech Republic}
\author{J. Kudrnovsk\'y}
\affiliation{Institute of Physics, Academy of Science of the Czech Rep.,
Na Slovance 2, Praha 8, CZ--18221}
\author{I. Turek}
\affiliation{Charles University,
Faculty of Mathematics and Physics, Department of
Condensed Matter Physics, Ke Karlovu 5, Praha 2, CZ--12116, Czech Republic}
\author{K. Uhl\'\i{}\v rov\'a}
\affiliation{Charles University,
Faculty of Mathematics and Physics, Department of
Condensed Matter Physics, Ke Karlovu 5, Praha 2, CZ--12116, Czech Republic}
\author{K. V\'yborn\'y}
\affiliation{Institute of Physics, Academy of Science of the Czech Rep.,
  Cukrovarnick\'a 10, Praha 6, CZ--16253}

\date{Mar20, 2020}   

\begin{abstract}
Temperature-dependent resistivity $\rho(T)$ and magnetoresistance are measured 
in bulk tetragonal phase of antiferromagnetic CuMnAs and the latter is
found to be anisotropic both due to structure and magnetic order. We
compare these findings to model calculations with chemical
disorder and finite-temperature phenomena included. The finite-temperature
{\em ab initio} calculations are based on the alloy analogy model
implemented within the coherent potential approximation and the
results are in fair agreement with experimental data. Regarding the anisotropic
magnetoresistance (AMR) which reaches a modest magnitude of 0.12\%, we 
phenomenologically employ the Stoner-Wohlfarth model to identify
temperature-dependent magnetic anisotropy of our samples and conclude
that the field-dependence of AMR is more similar to that of antiferromagnets
than ferromagnets, suggesting that the origin of AMR is {\em not}
related to isolated Mn magnetic moments.
\end{abstract}

\pacs{later}

\maketitle

The emergent field of antiferromagnetic (AFM) spintronics\cite{Baltz:2018_a}
has brought one particular AFM metal to prominence: CuMnAs. Apart from
electrical
switching\cite{Kaspar:2019_a} and domain wall manipulation\cite{Wadley:2018_a}
the main focus in exploring its response to electric field has so far been on
the optical range (ellipsometry and photoemission spectroscopy used to
validate band structure calculations\cite{Veis:2018_a}) and also on
the staggered spin polarisation induced by electric
field.\cite{Zelezny:2014_a} The latter led to the discovery of an
efficient means to manipulate\cite{Wadley:2016_a} magnetic moments in an AFM
and this, in turn, allowed the construction of memory
prototypes operating at room temperature\cite{Olejnik:2017_a} where
information is stored in the direction of magnetic moments. As a
method for read-out, anisotropic magnetoresistance (AMR) is used and
the primary aim of this work is to explore this very phenomenon in
CuMnAs. Contrary to previous recent studies of CuMnAs which entailed
epitaxially grown thin layers,\cite{Wadley:2013_a}
we now focus on bulk material.
 
\begin{figure}[b]
    \includegraphics[scale=0.35]{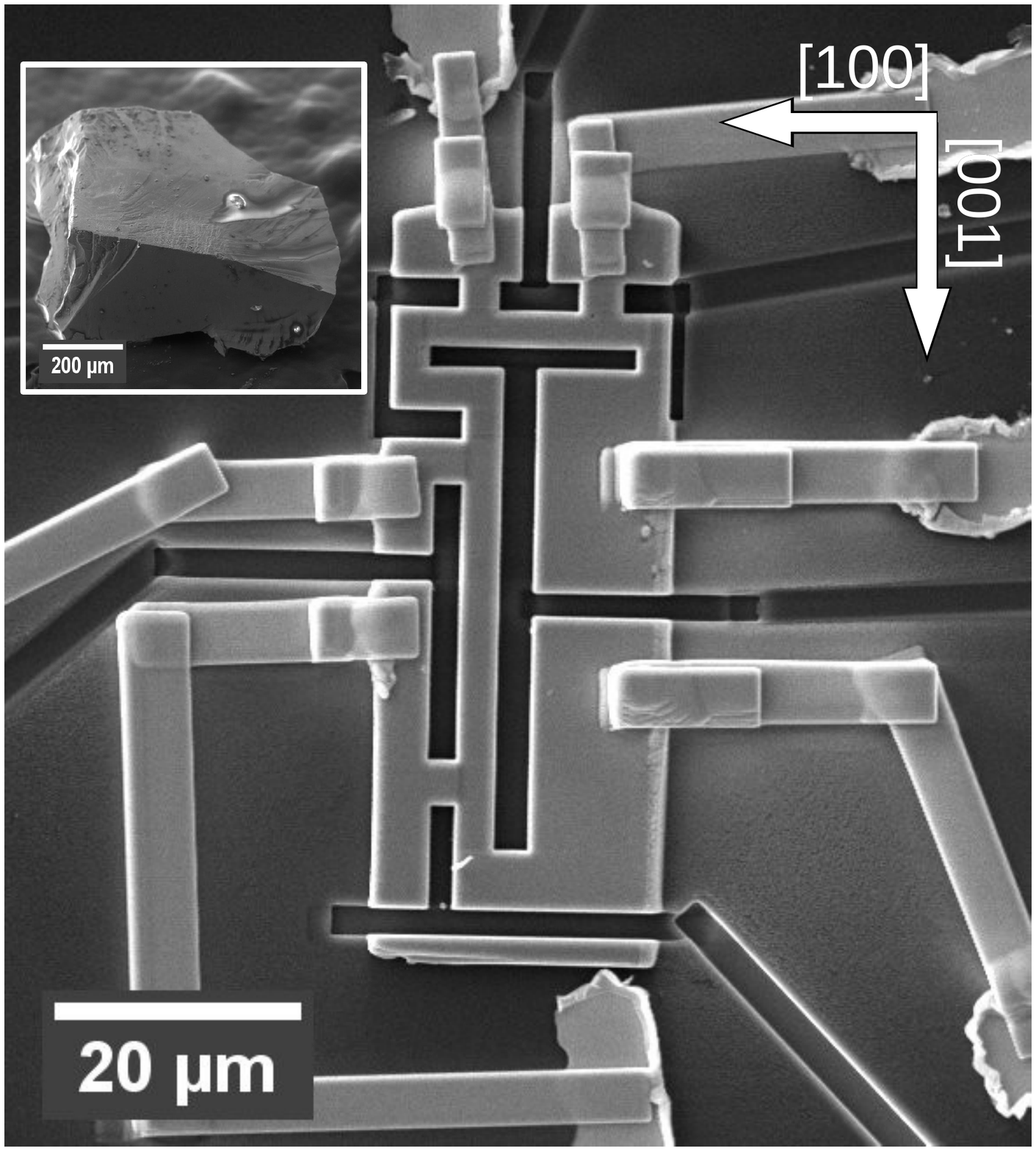}
\caption{Scanning electron microscope (SEM) image of the device and
  the single-crystalline grain from which it was fabricated. The [001]
  and [100] crystallographic directions correspond to the $c$ and $a$
  axes, respectively. Magnetic field $\vec{B}$ is rotated, with
  respect to this micrograph, from an in-plane direction
  $\vec{B}||[100]$ ($\psi=0$) to out-of-plane $\vec{B}||[010]$ ($\psi=\pi/2$);
  with respect to the crystallographic structure (see the inset in
  Fig.~\ref{fig-01}), $\vec{B}$ remains always in the basal plane ($a,b$).}
\label{fig-03}
\end{figure}

In the bulk form, CuMnAs was originally reported to have orthorhombic
structure\cite{Mundelein:1992_a} while thin films grown on GaP or GaAs
substrates\cite{Krizek:2019_a} adopt a tetragonal phase. Recent studies
of off-stoichiometric Cu$_{1+x}$Mn$_{1-x}$As compounds\cite{Uhlirova:2015_a}
have shown\cite{Emmanoulidou:2017_a,Uhlirova2019} that their
crystal structure is rather sensitive to the composition. While the
stoichiometric CuMnAs  compounds crystallise in an orthorhombic structure
(Pnma), few percent of copper excess at the expense of Mn turns the
structure to a tetragonal one (P4/nmm). In the tetragonal phase, the N\'eel
temperature reaches 507~K for Cu$_{1.02}$Mn$_{0.99}$As and decreases with
decreasing Mn content rather moderately; samples with more
off-stoichiometric composition have lower N\'eel temperatures, for
example $\approx 300$~K for\cite{Uhlirova2019} $x=0.4$. Our focus,
however, will be on the nearly-stoichiometric tetragonal systems.

The following section describes the fabrication of samples for
electrical transport measurements from a single-crystalline grain and
acquired experimental data. Modelling and interepretative approaches are
introduced in Sec.~II and sections III~and~IV are devoted to
models of zero-field transport and AMR, respectively. The two
appendices focus on magnetic anisotropy of CuMnAs and its modelling
and certain specialised aspects of microscopic transport calculations.

\section{Experimental}

\subsection{Growth and preparation}

A sample of tetragonal CuMnAs was prepared by reaction of high purity
copper, manganese and arsenic as previously reported.\cite{Uhlirova2019}
Tetragonal P4/nmm structure was confirmed by powder x-ray diffraction
at room temperature on Bruker D8 Advance diffractometer.\cite{Uhlirova2019}
Composition analysis performed by energy-dispersive x-ray detector
(EDX) suggests a slight prevalence of copper, the stoichiometry being
1.02(1):0.99(2):0.99(2) for Cu:Mn:As; N\'eel temperature ($T_N$) is 507~K.
From thus obtained polycrystal, a single-crystal grain was cleaved,
oriented using x-ray diffraction and its orientation was further
refined on an SEM stub holder using electron backscatter diffaction (EBSD).

For transport measurements, we adopted the sample fabrication
introduced by Moll~et~al.\cite{Moll2015,Moll2016,Moll2016:CdAs}
A rectangular lamella extending in the $ac$-directions
of dimensions 60$\times 20\times 3$~$\mu \mathrm{m}^{3}$ was
isolated out of a single-crystal grain using 30 kV Ga$^{2+}$ Focused Ion
Beam (FIB) Tescan Lyra XMH and transported to sapphire chip with
contact pads (5 nm Cr + 150 nm Au) prepared by photolithography. The
lamella was microstructured into a shape presented in Fig.~\ref{fig-03}
and it was conductively bonded to the contact pads using FIB assisted
chemical vapor deposition of Pt. Typical resistance of each contact
prepared by this method was around 50~$\Omega$. To improve the contact
resistance, we further sputtered the sample with a 100~nm Au~layer and
removed the excess gold from the top of our sample and in between the
contacts using the FIB. This resulted in an order of magnitude lower
resistance of 5~$\Omega$ per contact. 

%

This method allows us to precisely control the orientation of the
sample, which is essential due to highly anisotropic behaviour of
CuMnAs which we demonstrate in the following.  Furthermore, structuring
the sample into a long thin bar allows  us to obtain a high
signal-to-noise ratio without using high current and thus avoiding
self heating at low temperatures.

Resistivity measurement in a temperature range from 2 to 400 K was
carried out using a Quantum Design Physical Property Measurement System with
a Horizontal Rotator option. Typical currents were of the order of 100~$\mu$A
which translates into current densities ranging from $0.5$ to $2\times
10^7$~Am$^{-2}$ (small compared to what is used in CuMnAs-based memory
devices as writing  pulses\cite{Kaspar:2019_a}). The error in
calculating geometrical factor of the bulk device presented here is
about 15 \%.  This translates into a substantial part of error in
determining the bulk resistivity.

\begin{figure}
  \includegraphics[scale=0.45]{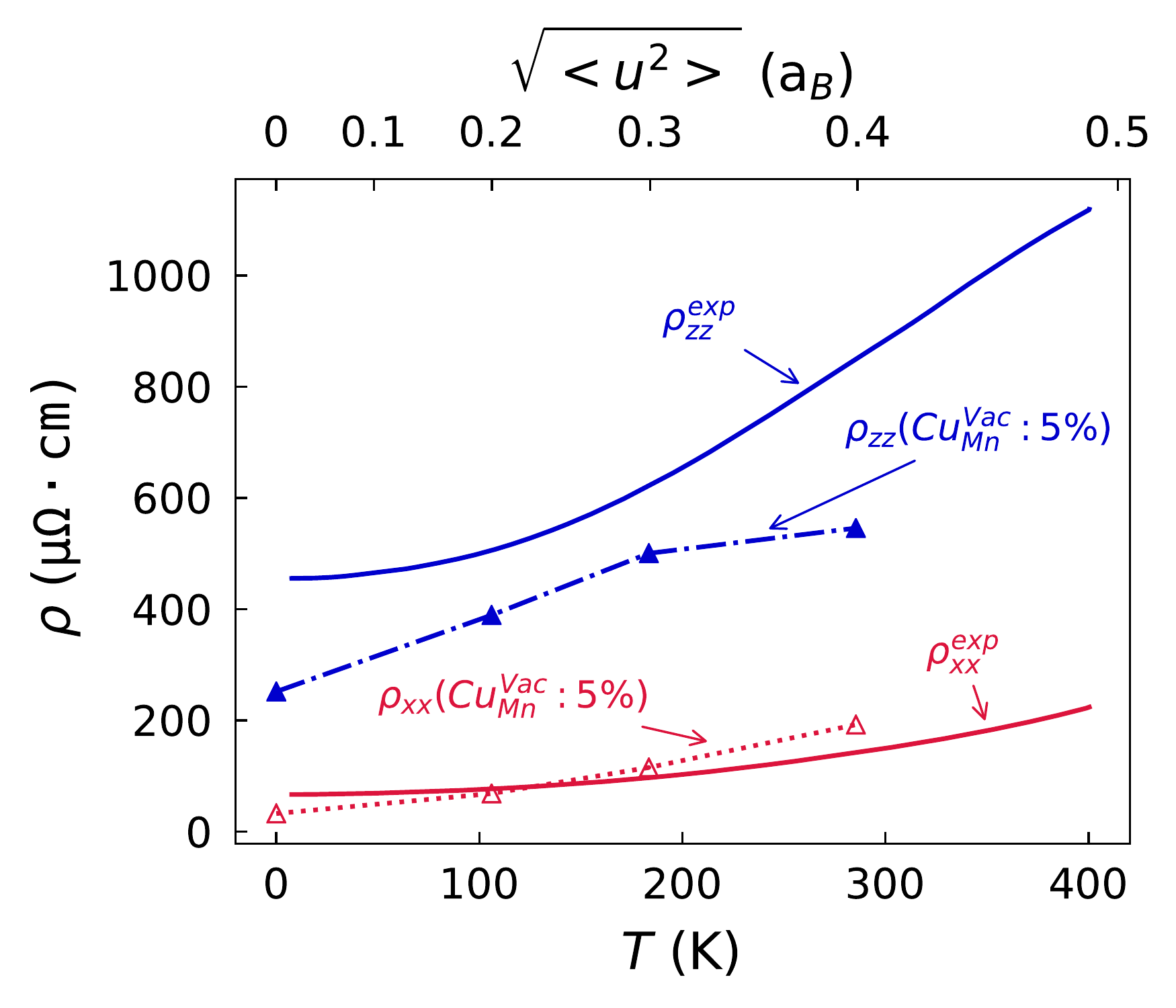}
\caption{Resistivity of bulk CuMnAs measured along $a$- (in-plane, $\rho_{xx}$)
  and $c$-axes (out-of-plane, $\rho_{zz}$) shown by solid lines;
  crystallographic axes are defined in the inset. To demonstrate
  a typical level of agreement with model calculations, resistivity
  assuming scattering on static impurities (Cu$_{\mathrm{Mn}}^{\mathrm{vac}}$
  as explained\cite{note3} later in Sec.~III) and phonons is also shown.}
  \label{fig-01}
\end{figure}

\subsection{Transport measurements}

Transport properties of tetragonal CuMnAs have previously been explored only in
thin films.\cite{Wadley:2013_a} Since all epitaxial growth processes
reported so far occur in the $(001)$ direction, only in-plane
resistivity ($\rho_{xx}$ or $a$-axis direction) can be found in
literature. Contrary to the thin films, our bulk devices allow for both
$\rho_{xx}$ and the out-of-plane component $\rho_{zz}$ to be measured
(here, we refer to crystallographic directions; both $\rho_{xx}$ and
$\rho_{zz}$ are measured in the plane of the lamella).
In-plane $\rho_{xx}$ data in Fig.~\ref{fig-01} are similar to previously
published results\cite{Wadley:2013_a} and we take notice of the large
structural anisotropy, i.e. resistivity along the $c$-axis being
almost an order of magnitude larger (at low temperatures, the ratio to
in-plane resistivity is $6.8\pm 0.8$ and it
slightly decreases at higher temperature). Given the layered structure of
CuMnAs, this fact is perhaps not very surprising. Low-temperature
$\rho_{xx}=67\pm10\ \mu\Omega\cdot$cm is somewhat lower (about 20\%) than
for thin layers of Ref.~\onlinecite{Wadley:2013_a}. This may be due to
slightly different composition of the compared materials or sample
quality; the residual resitivity ratios (RRRs) of bulk and thin films samples
are 2.2 and 1.8, respectively, and more recent samples\cite{Krizek:2019_a}
reach an even higher RRR of 3.
An example of model calculations in
Fig.~\ref{fig-01} is further discussed below (see Sec.~III): for now, the
data points (triangles) should only demonstrate the typical
level of agreement with one 
specific sort of impurities consistent with the known chemical composition of
the studied samples. We stress that a significantly better level of agreement
is achievable but only at the cost of less realistic model parameters (such
as impurity concentration).

\begin{figure}
  \begin{tabular}{ccc}
    \centering
    \includegraphics[scale=0.45]{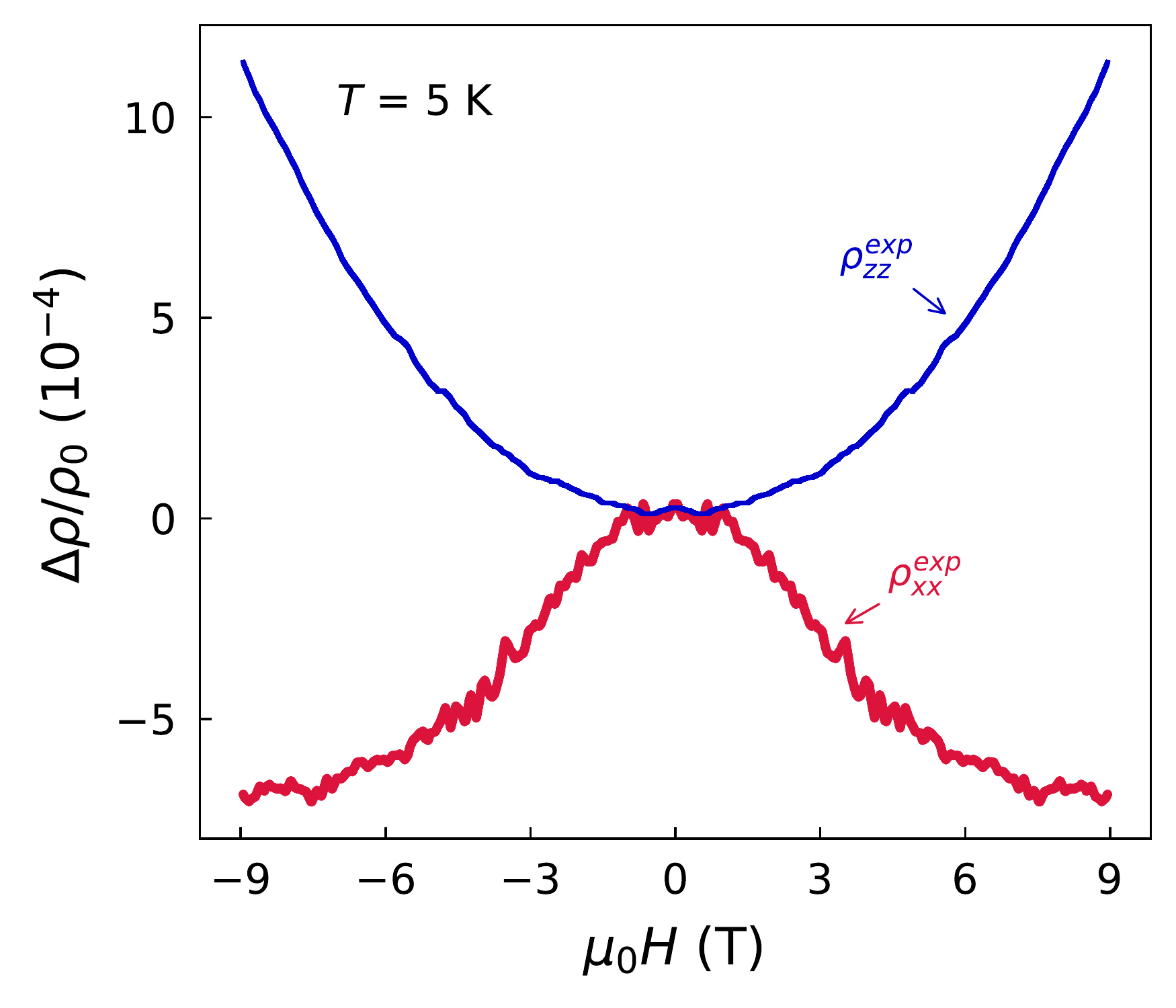}
  \end{tabular}  
  \caption{Low temperature magnetoresistance normalised to zero field
    value $R_0$ (which is different for the two directions of current).}
\label{fig-XX}
\end{figure}

When a magnetic field is switched on, we find a very different response
for in-plane and out-of-plane directions: the former shows a negative
magnetoresistance --- common in magnetic materials when an applied field
suppresses spin fluctuations --- but $\rho_{zz}(B)$ increases, see
Fig.~\ref{fig-XX}. In both cases, the magnetic field is perpendicular to
the current direction, i.e. along [010]. Apart from the AMR effect,
the negative magnetoresistance could be related to some kind of
magnetic moment response to the
applied magnetic field (it is prominent at lower $B$) while the usual
positive magnetoresistance in metals dominates at larger magnetic
fields. For current along the $c$-axis, the manipulation of magnetic
moments is of no effect (they always remain perpendicular to the
current direction) and only the positive magnetoresistance remains.

\begin{figure} 
  \begin{tabular}{c}
    \includegraphics[scale=0.35]{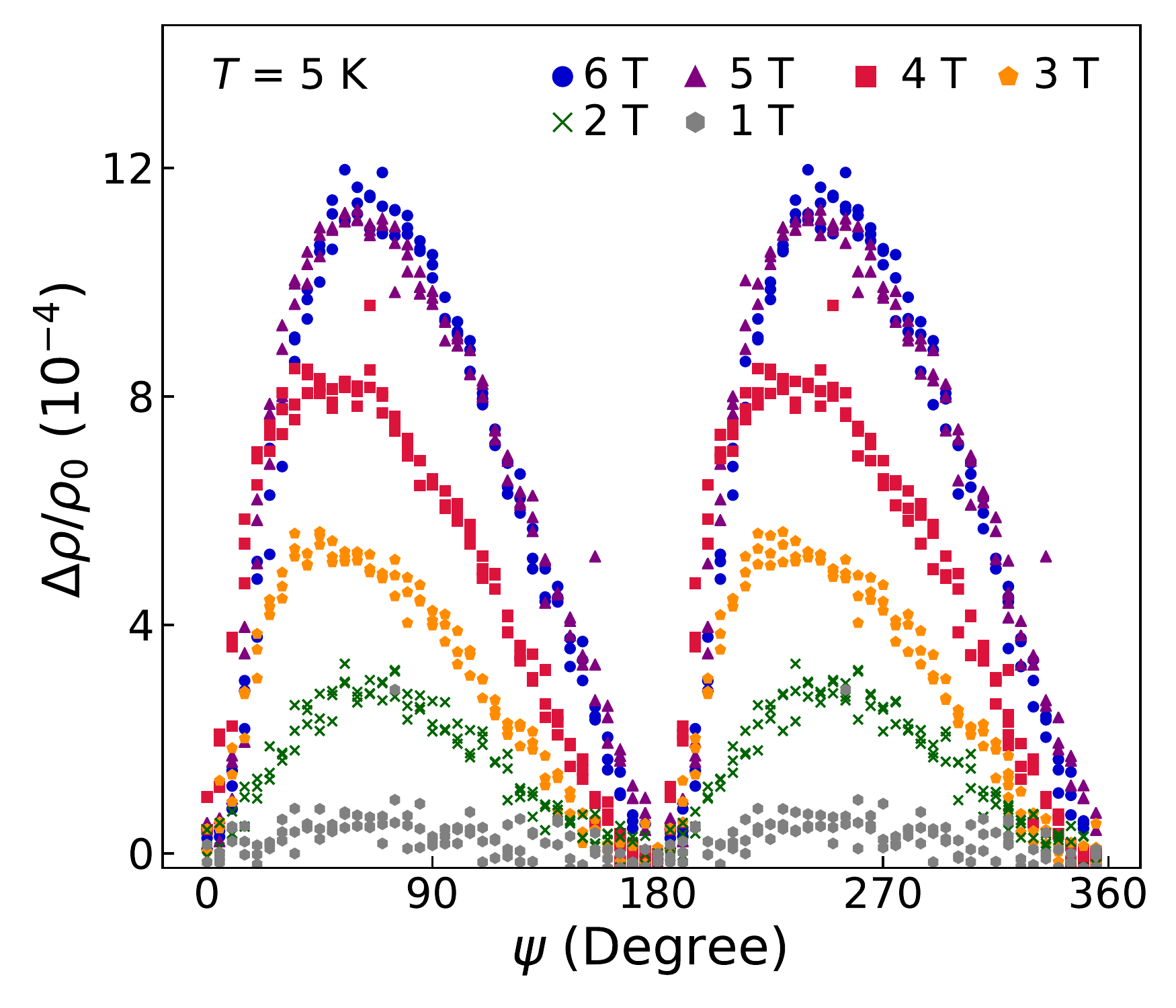} \\
    \includegraphics[scale=0.35]{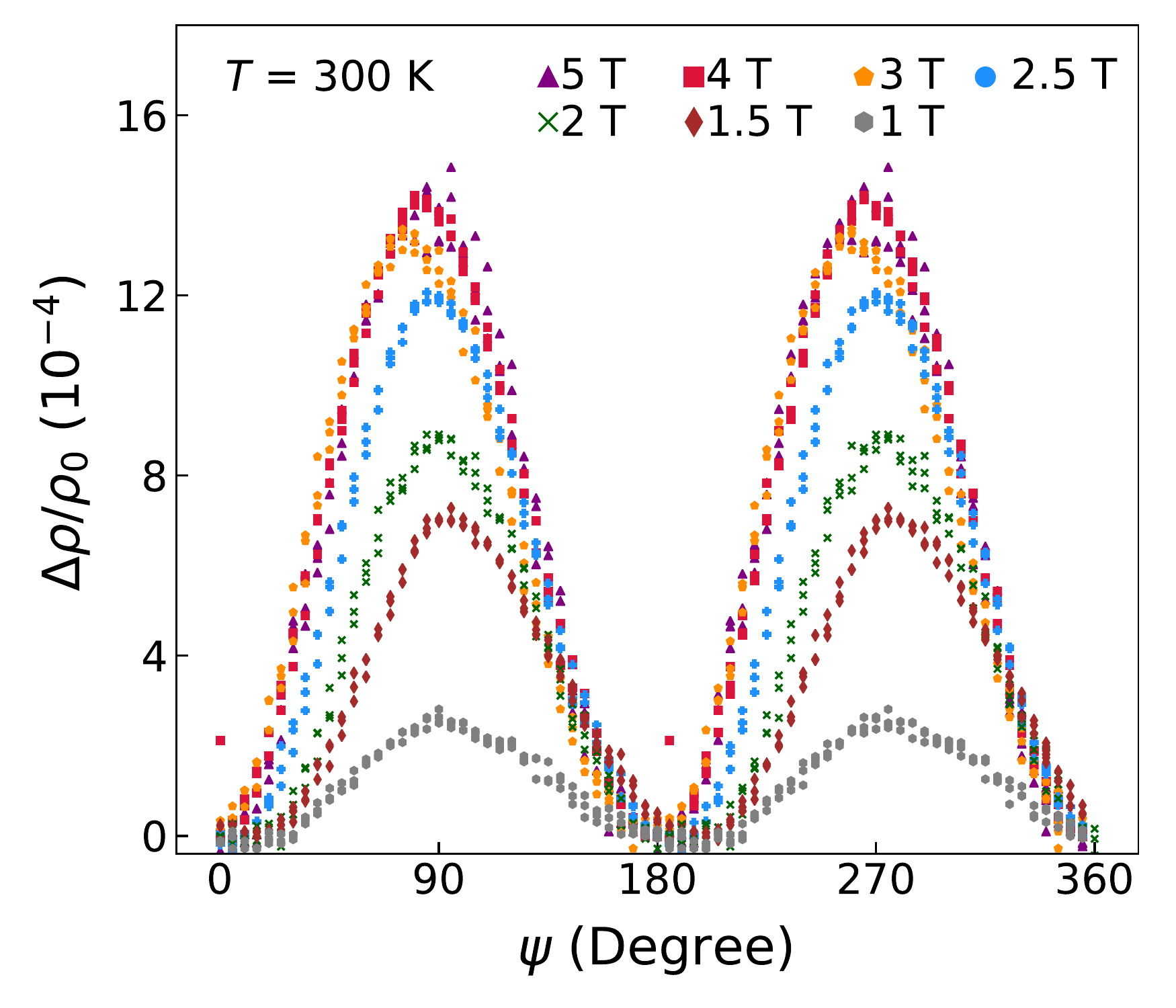} \\ 
    \includegraphics[scale=0.35]{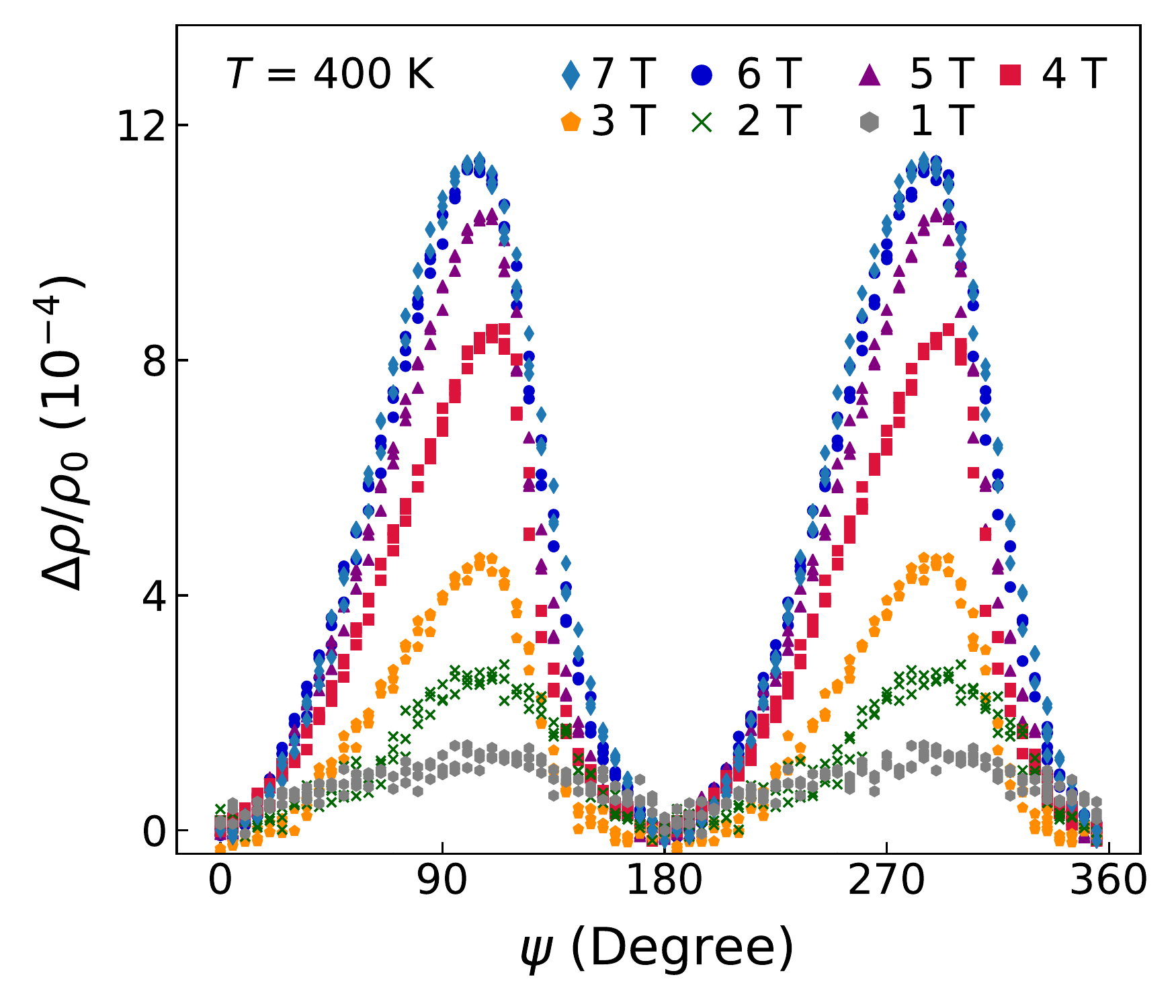} 
  \end{tabular}
\caption{Angular sweeps (magnetic field rotated in the plane) of $\rho_{xx}$
showing AMR which deviates from the $\Delta\rho_{xx}\propto \cos^2\psi$
dependence. Within the sweep, $\rho_0$ is the minimum value of
resistivity. {\em Top:} low temperature, {\em bottom:} high
temperature, {\em middle:} intermediate temperature where the
deviation is suppressed.}
\label{fig-02}
\end{figure}

Focusing on in-plane magnetotransport, we also find a clear anisotropy
(i.e. $\rho_{xx}$ different from $\rho_{yy}$ when
$\vec{B}\parallel\hat{x}$).\cite{note1}   Here, it should be noted
that large magnetic anisotropies force the N\'eel vector
into the $ab$-plane (see Appendix~A) and conceivably, there remain
weak in-plane anisotropies which allow for the magnetic moments to be 
moved within the plane easily. Angular
sweeps shown in Fig.~\ref{fig-02} suggest both the presence of AMR and
temperature-dependent magnetic anisotropies which we discuss in 
Sec.~IV. We observe a gradual increase of the AMR amplitude up to
$\approx 6$~T and above this magnetic field, the AMR signal does not
change (measured up to 9 T, not shown). Low temperature ($T=4$~K) and
close-to-N\'eel-temperature ($T=400$~K~$<T_N$) measurements show clearly
different distorsion of the $\Delta\rho_{xx}(\psi)\propto\cos 2\psi$
signal, see also Eq.~(\ref{eq-05}). Such cosine-squared form would be
typical of polycrystalline samples\cite{Rushforth:2007_a} if
magnetocrystalline anisotropy were negligible ($\psi$ is the angle
between $\vec{B}$ and the current direction, see Fig.~1). 

\section{Introduction to modelling}

We employ two approaches to interpret our measured data: a microscopic
model of electric transport where the direction of magnetic moments
present in the system is assumed to be known; and a phenomenological
one based on a Stoner-Wohlfarth model where the coupling between
external magnetic field and magnetic order of CuMnAs samples is
investigated. The latter approach allows to partially overcome our
lack of knowledge about the precise nature of potential magnetic
impurities. It serves the purpose of interpreting angular sweeps in
Fig.~\ref{fig-02} where the externally controlled parameter is $\psi$
rather than directly the magnetic moments.

Our microscopic modelling is based on the tight-binding linear muffin-tin 
orbital (TB-LMTO) method with the atomic sphere approximation and the
multicomponent coherent potential approximation (CPA) \cite{IT-book}.
Calculations employ the Vosko-Wilk-Nusair exchange-correlation
potential\cite{Vosko1980} and Hubbard $U$ was used in the fully
relativistic LSDA+$U$ scheme for $d$-orbitals of Mn, similarly to
Refs.~\onlinecite{DW2019-JMMM} (TB-LMTO) and \onlinecite{Veis:2018_a}
(LAPW). The value $U=0.1$~Ry quoted in
Tab.~\ref{tab-01} was found consistent with optical and photoemission
spectra in the latter reference. The scalar-relativistic methods
(see Tab.~\ref{tab-03} in Appendix~B) are used only for a comparison with
previous results.\cite{Maca:2017_a,Maca2019}  Band structures yielded by
different approaches (including $GW$) can be found in
Ref.~\onlinecite{DW2019-JMMM}.

Electrical transport properties are studied in a framework of the
linear response theory and the Kubo-Bastin formula,\cite{Turek:2012_a}
the velocity operators describe intersite
hoppings\cite{IT-transport} and we take into account CPA-vertex
corrections.\cite{KC-multilayers} 
Longitudinal conductivities are given only by the Fermi-surface term;
therefore, the Fermi-sea contribution \cite{IT-FermiSea} is omitted.  
Finite-temperature atomic displacements (phonons) are treated by alloy analogy
model (AAM)\cite{Ebert2015,Kodderitzsch2013,Glasbrenner2014,Starikov2018};
this model has recently been incorporated into the TB-LMTO-CPA technique.

For the inclusion of phonons, an extended \textit{spdf-}basis is needed
because of
transformations of the LMTO potential functions.\cite{DW2017-IEEE, DW2017-SPIE}
To compare novel results with literature,\cite{Maca:2017_a,Maca2019} a few
\textit{spd-}calculations are shown in Appendix (Tab.~\ref{tab-03}).
Fluctuations of magnetic moments at nonzero temperatures are included
by the tilting model\cite{DW-newJMMM} which was shown to describe
low-temperature electrical transport of CuMnAs well.\cite{DW2019-JMMM}
Fluctuations of magnetic moments at nonzero temperatures are included
only by the disordered local moment (DLM) approach.\cite{Kudrnovsky:2012_a}
Tilting of magnetic moments from their equilibrium direction could be
also included within the AAM\cite{Starikov2018} as well as our TB-LMTO
AAM\cite{DW2019-JMMM}, but it is beyond the scope of this study.
Zero-temperature calculations that involve magnetic impurities (such
as Mn atom substituting Cu or As) are also based on the DLM approach.
With this machinery at hand, temperature-dependent resistivity can
successfully be modelled, provided we specify the source of
scattering at $T=0$ (otherwise, $\rho\to 0$ at low
temperatures).


\begin{table}[h]
  \begin{tabular}{|c|c|cc|cc|}\hline
        &       Formation       &       \multicolumn{4}{c|}{Resistivity [$\mu\Omega$cm]}                                                  \\
        &       energy  &       \multicolumn{2}{c|}{$U=0.00\,\text{Ry}$}                        &       \multicolumn{2}{c|}{$U=0.10\,\text{Ry}$}                        \\
Defect  &       [eV] \cite{Maca2019}    &       $\rho_{xx}$     &       $\rho_{zz}$     &       $\rho_{xx}$     &       $\rho_{zz}$     \\ \hline
Vac$_\textrm{Mn}$ &     $-0.16$   &       31      &       184     &       20      &       181     \\
Vac$_\textrm{Cu}$ &     $-0.14$   &       16      &       79      &       11      &       92      \\
Mn$_\textrm{Cu}$  &     $-0.03$   &       112     &       263     &       150     &       915     \\
Cu$_\textrm{Mn}$  &       0.34    &       23      &       131     &       8       &       57      \\
Cu$_\textrm{As}$  &       1.15    &       121     &       481     &       163     &       1299    \\
As$_\textrm{Cu}$  &       1.73    &       114     &       359     &       123     &       694     \\
As$_\textrm{Mn}$  &       1.79    &       141     &       476     &       161     &       617     \\
Mn$_\textrm{As}$  &       1.92    &       147     &       423     &       186     &       1784    \\
Vac$_\textrm{As}$ &       2.18    &       210     &       306     &       284     &       1556    \\
Cu$\leftrightarrow$Mn   &       -       &       120     &       393     &       142     &       882     \\ \hline
  \end{tabular}  
  \caption{Comparison of various impurity types in tetragonal CuMnAs
    (e.g. Vac$_\textrm{Cu}$ or Mn$_\textrm{Cu}$ indicate a copper
    vacancy and Mn atom substituting Cu, respectively).    Calculated
    formation energy suggests that impurities involving arsenic are
    unlikely. Fully relativistic $spdf$ calculations of resistivity are
    given for 5\% of the respective impurity.}
  \label{tab-01}
\end{table}

\section{Ab initio transport at zero field}

We first focus on residual resistivity. Experimentally, we know that
stoichiometry of our CuMnAs samples is 1:1:1 within a few per cent margin and
that puts a limit of maximum impurity concentration. Tab.~\ref{tab-01}
gives an overview of  calculated resistivities for various types of
impurities (5\% of the respective impurity).
It should be pointed out that fundamentally different sources of
scattering than those listed in Tab.~\ref{tab-01} may occur (even at
zero temperature), e.g. structural defects such as 
linear dislocations
have recently\cite{Krizek:2019_a} been identified in epitaxial layers.

With this provision, the following conclusions can be drawn regarding
resistivities in the absence of external magnetic field. (i) In a very
broad picture, all of the listed values of resistivities are
plausible; note that exact concentration of impurities is not known
for our samples so even large values of $\rho$ seen in Tab.~\ref{tab-01}
could be compatible with experimental data in Fig.~\ref{fig-01}
supposing the given type of impurity occurs at a low concentration.
(ii) All listed cases involve a clear structural anisotropy
$\rho_{zz}/\rho_{xx}>1$.   These two basic observations do not
principially exclude any of the options in the table, however,
(iii) defects involving arsenic, both as a dopant or as a site to be
occupied by another atom (substitutional or interstitial positions), seem
unlikely given prohibitive formation energies.\cite{Maca2019}   (iv) Among
the five remaining options, those compatible with Cu-rich
stoichiometry show resistivity somewhat low compared to experimental data.
(v) At this point (i.e. based on calculations in Tab.~\ref{tab-01}),
the most likely scenario, disregarding additional sources of
scattering,\cite{Krizek:2019_a} would thus entail a combination of at
least  two types of impurities: for example Cu substituting Mn
(Cu$_{\mathrm{Mn}}$) and a Cu/Mn swap (Cu$\leftrightarrow$Mn), both at
concentrations of few per cent.

\begin{figure}
\includegraphics[scale=0.5]{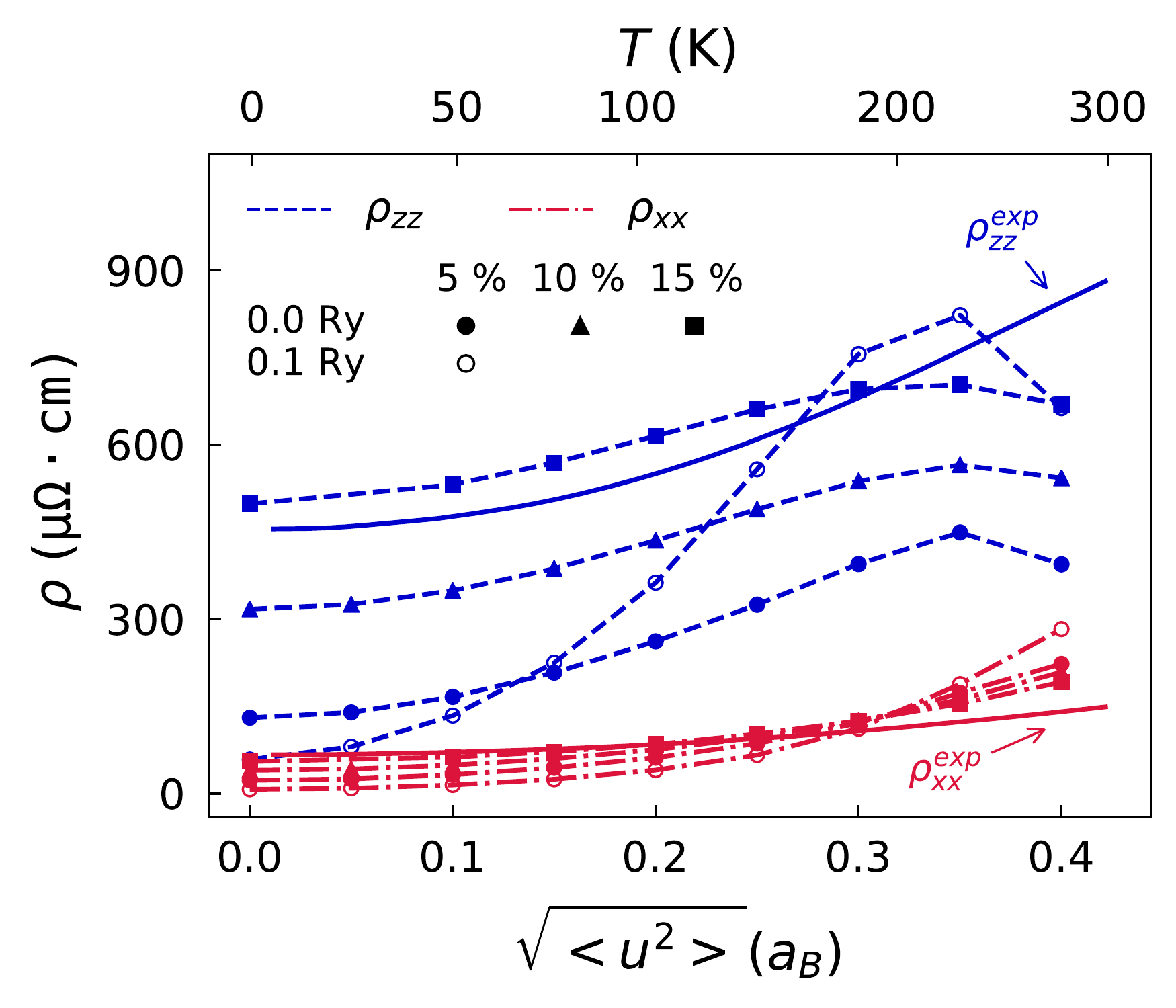}
\caption{Resistivity of CuMnAs calculated microscopically assuming
  finite-temperature atomic
  displacements $\sqrt{\langle u^2\rangle}$ (modelling phonons)
  and Cu$_{\mathrm{Mn}}$ impurities
  (modelling the source of residual resistance at low temperature)
  at various concentrations.}
\label{fig-08}
\end{figure}

Next, we consider the temperature dependence of resistivity and here,
the primary source of scattering are the phonons. As a note of caution,
we remark that calculated results are plotted as a function of
$\sqrt{\langle u^2\rangle}$ and conversion\cite{DW2019-PRB} into $T$ requires
the knowledge of Debye temperature $T_D$. (We use the value from
orthorhombic phase, see Ref.~\onlinecite{DW2019-JMMM} for
explanation.)  Calculations with 5\% of Cu$_{\mathrm{Mn}}$ and $U=0$
in Fig.~\ref{fig-08} show a reasonable trend but overall values (in
particular, of $\rho_{zz}$) are too low. Combination with other types
of impurities such as\cite{note3} Cu$_{\mathrm{Mn}}^{\mathrm{vac}}$
offers a partial remedy (see model data in Fig.~\ref{fig-01})
but since concentration-dependence of resistivity is not always linear
(see Appendix~B and Fig.~\ref{fig-05}), construction of a quantitative
model is difficult.
We note that decreasing
resistivity for high magnitudes of atomic displacements 
(see Fig.~\ref{fig-08}) is probably caused by an increase of
DOS at the Fermi level, similarly  the effect of magnons and
phonons.\cite{DW2019-JMMM}  The same effect may be responsible for
nonmonotonic dependence of resistivity on concentration of
Cu$_{\mathrm{Mn}}^{\mathrm{vac}}$: both observations clearly
contradict the Matthiessen rule and are further discussed in Appendix~B.   

With temperature--dependent resistivity, phonons are not the sole
source of scattering to be considered; rather, combined effect of
impurities, phonons, and magnons should be taken into account. 
Above, we have shown a deviation of the resistivity from Matthiessen's
rule for impurities and phonons; in Ref.~\onlinecite{DW2019-JMMM}, the
same was reported for phonons and magnons. In that reference,
we numerically justified a collinear uncompensated disordered local
moment (uDLM) model of spin fluctuations and we demonstrated, that the
tilting model of the magnetic disorder agrees well with experimental
data up to room temperature. We now adopt the second approach and
illustrate the combined effect of phonons and
magnons and static Cu$_{\mathrm{Mn}}$ impurities. A similar model was
discussed in Ref.~\onlinecite{Kelly2018} (relativistic effects in this
context can also be considered\cite{twoRefsDavid}).
A decrease of mean local magnetic moment of Mn atoms was
mapped on Monte-Carlo simulations\cite{Maca2019} to obtain the temperature
dependence of the spin fluctuations.\cite{DW2019-JMMM,DW2019-PRB}
Data presented in Tab.~\ref{tab-02} show that even for lower concentration
of Cu$_{\mathrm{Mn}}$, the combined effect of phonon and magnon
scattering close to the room temperature leads to $\rho_{xx}$ clearly
exceeding the experimental values while $\rho_{zz}$ remains
underestimated.

\begin{table}[h]
\caption{Resistivity (in $\mu\Omega$ cm) due to a combined
  effect of static impurities (the sole source of scattering at
  $T=0$), phonons and magnons (tilting model).}
  \begin{tabular}{|c|l||cc|cc|}
\hline    
$T$&& Cu$_{\rm{Mn}}$: 5 \%&& Cu$_{\rm{Mn}}$: 10 \%&\\
$\mathrm{[K]}$     &       Effects  &       $\rho_{xx}$     &       $\rho_{zz}$     &       $\rho_{xx}$     &       $\rho_{zz}$     \\ \hline\hline
0       &       -       &       23      &       131     &       41      &       319     \\ \hline
        &       Ph.     &       39      &       190     &       55      &       371     \\
65      &       Mag.    &       40      &       215     &       55      &       428     \\
        &       P.+M.   &       59      &       269     &       72      &       474     \\ \hline
        &       Ph.     &       172     &       450     &       161     &       566     \\
230     &       Mag.    &       115     &       474     &       110     &       724     \\
        &       P.+M.   &       257     &       345     &       263     &       450     \\ \hline
  \end{tabular}
  \label{tab-02}
\end{table}

The underestimated values of structural anisotropy
$\rho_{zz}/\rho_{xx}$ seem to be a general feature of our calculations.
%
Previous calculations were obtained without $U$ (except for one
dataset in Fig.~\ref{fig-08} which we wish to discuss now); however,
the electronic structure has not yet been reliably determined and
the LSDA+$U$ agrees the best with $GW$ calculations\cite{DW2019-JMMM} 
when $U=0.20$~Ry. We emphasize, that the band strucutre
pertains to CuMnAs without any disorder, while the transport is
studied in disordered samples.  Therefore, we consider the band
strucutre to be of lesser importance for explaining the electrical
transport than the DOS. The temperature-dependent resistivity already for
$U=0.10$~Ry increases about twice faster than both measured data and
$U=0.00$~Ry calculations; see Fig.~\ref{fig-08} for
Cu$_{\mathrm{Mn}}$. We have investigated also the role of  $U$ on
other impurities and
finite-temperature disorder (not shown here) and, in general, nonzero
Hubbard parameter makes both increase and decrease of the resistivity
more significant (compared to $U$ vanishing). This can be attributed to
decreasing DOS\cite{DW2019-JMMM} around $E_F$ for increasing $U$ and,
therefore, a larger sensitivity of electrical transport on small changes
(caused by impurities or finite-temperature disorder).

\section{AMR Modelling}

Experimental data (angular sweeps in Fig.~\ref{fig-02}) show a
pronounced AMR with two-fold symmetry reaching $\Delta
\rho_{xx}/\rho_0\sim 10^{-3}$ at saturation (here, $\rho_0$ is the
planar average of resistivity). Theoretical data, see 
Tab.~\ref{tab-04}, suggest that this magnitude of AMR is compatible
with basically any type of static disorder considered so far. Larger
theoretical values (compared to the measured ones), however, indicate
that it is not the whole system that responds to the applied magnetic
field $\vec{B}$: for example, magnetic anisotropy for a large part of
magnetic moments would not be overcome by $\vec{B}$ available in our
experiments and only few free moments would move. (Such free moments
could be related to structural defects.)
In the case of Cu/Mn swaps, the difference is extreme so either
this defect is not very common in our samples or it is largely
insensitive to $\vec{B}$.

\begin{table}[h]
  \begin{tabular}{|c|c|cc|cc|}\hline
        &       \multicolumn{2}{c|}{Fully rel., \textit{spd}}                   &       \multicolumn{2}{c}{Fully rel., \textit{spdf}}     &             \\
Defect  &       $U=0$     &       $U=0.10\,\mbox{Ry}$     &       $U=0$     &       $U=0.10\,\mbox{Ry}$   &  \\ \hline
Vac$_\textrm{Mn}$ &       6.09$\cdot 10^{-3}$     &       1.16$\cdot 10^{-2}$     &       -2.08$\cdot 10^{-4}$    &       2.01$\cdot 10^{-2}$   &  \\
Vac$_\textrm{Cu}$ &       -1.04$\cdot 10^{-2}$    &       1.08$\cdot 10^{-2}$     &       5.24$\cdot 10^{-3}$     &       -1.85$\cdot 10^{-2}$  &  \\
Mn$_\textrm{Cu}$  &       2.52$\cdot 10^{-3}$     &       6.25$\cdot 10^{-4}$     &       2.29$\cdot 10^{-3}$     &       1.59$\cdot 10^{-3}$   &  \\
Cu$_\textrm{Mn}$  &       6.69$\cdot 10^{-3}$     &       -5.32$\cdot 10^{-4}$    &       -2.05$\cdot 10^{-3}$    &       1.34$\cdot 10^{-2}$   &  \\
Cu$_\textrm{As}$  &       1.70$\cdot 10^{-3}$     &       1.03$\cdot 10^{-3}$     &       1.66$\cdot 10^{-3}$     &       1.80$\cdot 10^{-4}$   &  \\
As$_\textrm{Cu}$  &       2.79$\cdot 10^{-3}$     &       1.05$\cdot 10^{-3}$     &       2.42$\cdot 10^{-3}$     &       1.13$\cdot 10^{-3}$   &  \\
As$_\textrm{Mn}$  &       3.41$\cdot 10^{-3}$     &       1.31$\cdot 10^{-3}$     &       1.60$\cdot 10^{-3}$     &       1.30$\cdot 10^{-3}$   &  \\
Mn$_\textrm{As}$  &       2.95$\cdot 10^{-3}$     &       1.20$\cdot 10^{-3}$     &       2.47$\cdot 10^{-3}$     &       9.98$\cdot 10^{-4}$   &  \\
Vac$_\textrm{As}$ &       2.17$\cdot 10^{-3}$     &       1.87$\cdot 10^{-5}$     &       2.99$\cdot 10^{-3}$     &       2.27$\cdot 10^{-3}$   &  \\
Cu$\leftrightarrow$Mn   &       2.54$\cdot 10^{-1}$     &       2.03$\cdot 10^{-1}$     &       2.14$\cdot 10^{-1}$     &       1.16$\cdot 10^{-1}$   &  \\ \hline
  \end{tabular}
  \caption{Theoretical AMR, i.e. $(\rho_{xx}-\rho_{yy})/\rho_{av}$,
    resulting from the microscopical model and
    $\rho_{av}=(\rho_{xx}+\rho_{yy})/2$.
    Calculations assume 5\% of
    the respective impurity and magnetic moments along $x$. }
  \label{tab-04}
\end{table}

Without specifying what in reality responds to magnetic field (bulk of
the system, decoupled magnetic moments etc.), we can phenomenologically
use the Stoner-Wohlfarth (SW) model to analyse data in Fig.~\ref{fig-02}. It
can easily be adapted to study either ferromagnets (as originally
conceived\cite{SW48}) or antiferromagnets\cite{Correa:2018_a}. In the latter
case, energy (per volume) divided by sublattice magnetisation $M$ reads 
\begin{equation}\label{eq-01}
  \frac{E}{MV}= B_e\vec{m}_1\cdot\vec{m}_2 -B\vec{b}\cdot(\vec{m}_1+\vec{m}_2)
  +B_a [(\vec{m}_1\cdot \hat{a})^2+(\vec{m}_2\cdot \hat{a})^2].
\end{equation}
while for ferromagnets, the exchange term (described by field $B_e$)
between sublattices $\vec{m}_{1,2}$ is not present
\begin{equation}\label{eq-02}
  \frac{E}{MV}= -B\vec{b}\cdot \vec{m} +B_a (\vec{m}\cdot \hat{a})^2
\end{equation}
and only a single magnetic moment direction $\vec{m}$ is considered (all
$\vec{m}_{1,2}$, $\vec{m}$ and $\vec{b}=\vec{B}/B$ are unit vectors).
Magnetic anisotropy (see Appendix~A) is assumed to have a uniaxial form
(the axis being a general in-plane direction $\hat{a}$)
and it is represented by field $B_a$. Minimising the
energy given by Eqs.~\ref{eq-01} or~\ref{eq-02}, the direction of 
$\vec{m}_{1,2}$ (or $\vec{m}$) can be determined for arbitrary
direction and magnitude of $\vec{B}$. Assuming that the AMR is
dominated by non-crystalline terms\cite{Rushforth:2007_a}
\begin{equation}\label{eq-05}
  \frac{\Delta \rho_{xx}(\phi)}{\rho_0} = C_I \cos 2\phi
\end{equation}
the angular sweeps in $\psi$, $\vec{b}\cdot\hat{x}=\cos\psi$ can be
simulated. The SW model provides the connection, via energy minimisation,
between $\psi$ (as an input) and $\phi$ (as an output) which is the angle
between current direction and $\vec{m}$ or N\'eel vector.

As a matter of fact, the SW model for both ferromagnet (FM) and
antiferromagnet (AFM), Eqs.~\ref{eq-01},\ref{eq-02}, reduces to
{\it almost} the same form if $\vec{m}_{1,2}$ are assumed to lie in
plane so that their direction can be represented by a single angle $\phi$:
\begin{equation}\label{eq-03}
  \tilde{E} = 2\alpha \cos 2\phi - \frac12 \beta^n \sin^n(\phi-\psi),
\end{equation}
where $n=1$ for a FM and $n=2$ for an AFM and $\psi$ represents the
direction of the magnetic field. Particular expressions for $\alpha$
and $\beta$ differ for the FM and AFM flavours of the model but in both
cases, $\alpha\propto B_a$ relates to the magnetic anisotropy and
$\beta\propto B$ describes the effect of external magnetic field; see
Appendix~A for detailed explanation. We stress that attempts to model
the data with biaxial anisotropy (which would be more natural in a
tetragonal system) lead to visibly worse quality of fits.

For practical purposes, the difference between $\sin$ and $\sin^2$ is
unimportant in modelling results: in both cases, the second term in
Eq.~(\ref{eq-03}) provides a minimum close to $\phi=\psi$. The only
substantial difference between the FM and AFM cases is, effectively, how the
Zeeman-like term depends on magnetic field ($\propto \beta^2$ for AFM,
$\propto\beta$ for FM) and this allows for a straightforward test of
experimental data. We first fit the measured data at $B$ large enough
for saturation, see Fig.~\ref{fig-04}, and determine $\alpha$ in
Eq.~(\ref{eq-03}) assuming that $\beta=1$. In Appendix~A, we explain
the fitting procedure in detail and here we only remark that the
effective magnetic anisotropy implied by angular sweeps data
does not have the easy axis $\vec{a}$ aligned with any high symmetry
direction. Measurements at different temperatures $T$ are consistent with
$\vec{a}$ being independent of $T$ whereas the magnitude of the
anisotropy $\propto B_a$ does change and even flips the sign. This
is manifest in different shapes of data on the left and right
panels of Fig.~\ref{fig-04}.

\begin{figure}
\includegraphics[scale=0.23]{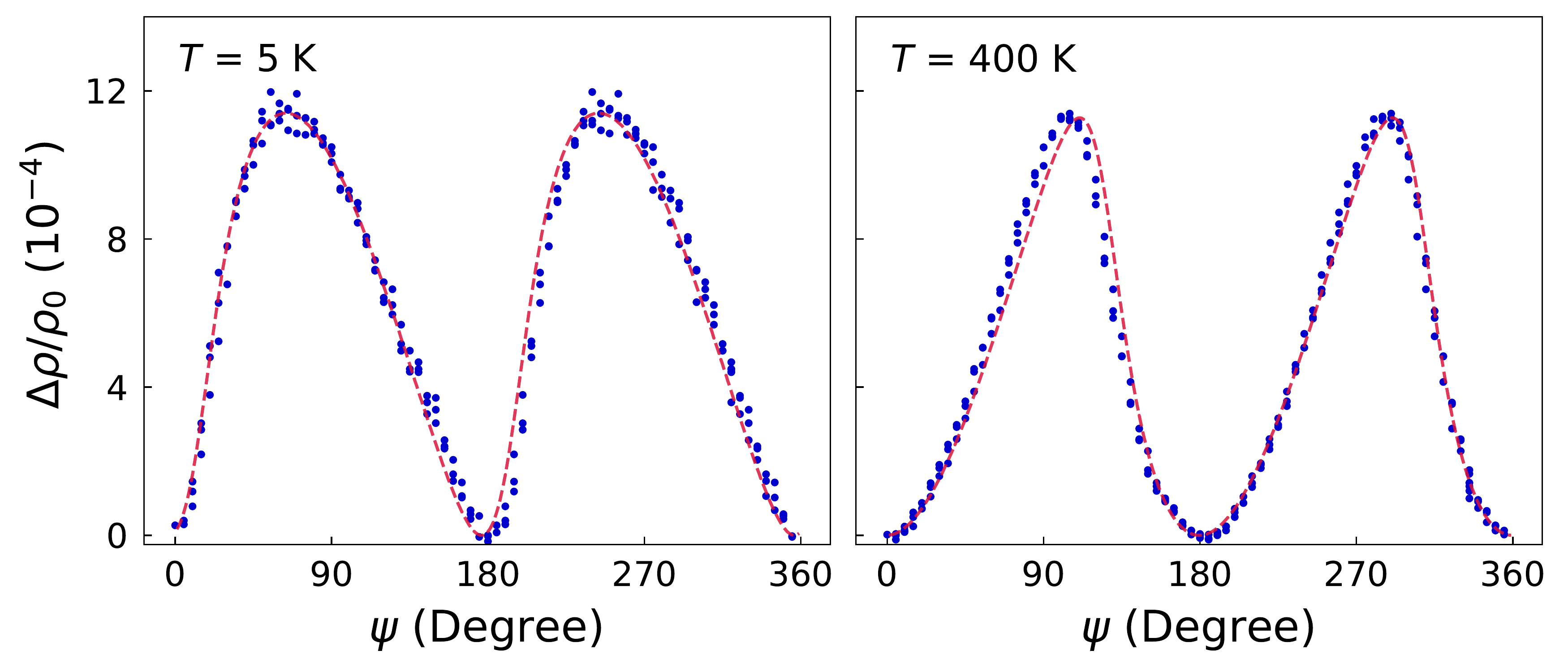}
  \caption{SW analysis of angular sweeps at maximum $B$ (left/right:
  low/high temperature). Data taken from Fig.~\ref{fig-02}.}
\label{fig-04}
\end{figure}

\begin{figure}
\includegraphics[scale=0.23]{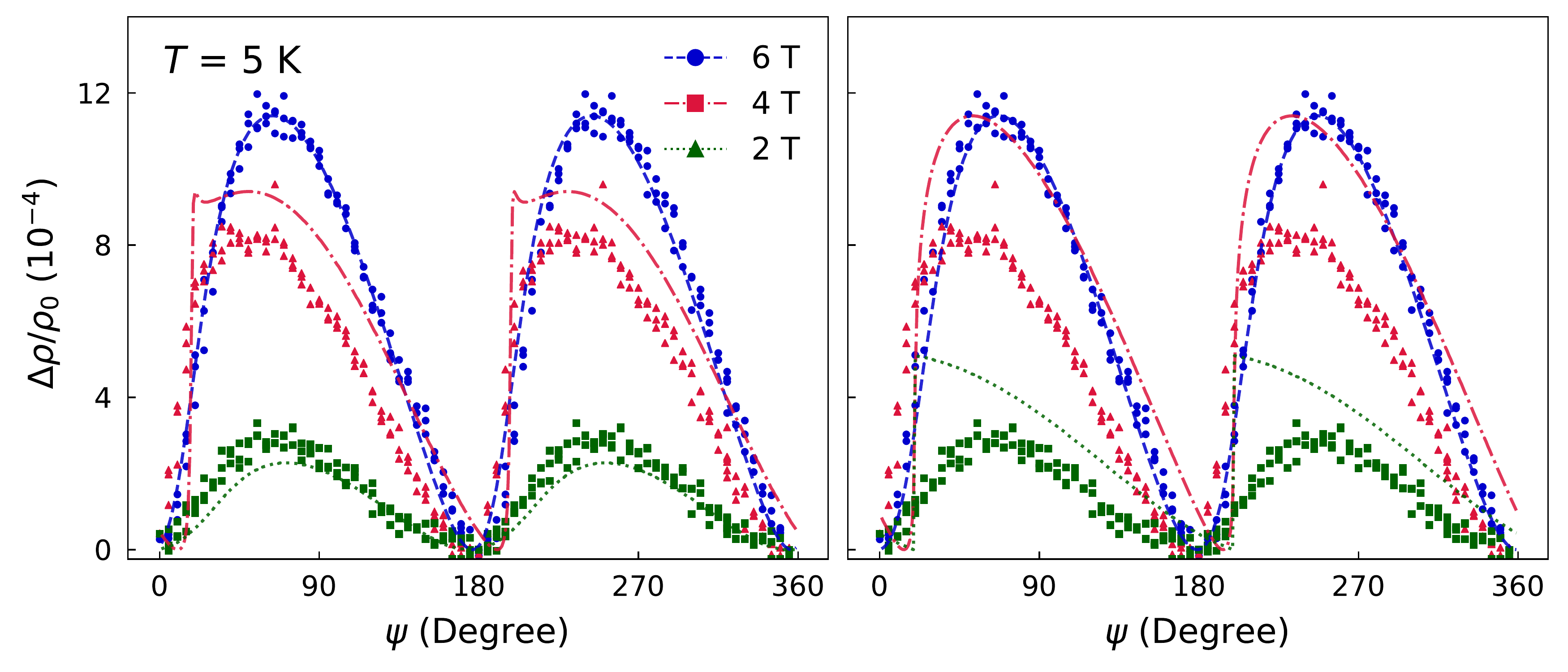}
\caption{
    Analysis of the field-dependence ($T=5$~K data from Fig.~\ref{fig-02})
    based on two flavors of SW model: (a) AFM, (b) FM. }
\label{fig-06}
\end{figure}

Next, we use the fitted parameters ($B_a$ from Eq.~\ref{eq-01}) and
look at lower $B$ than the saturation field: the FM model (drop the
$B_e$ term in Eq.~\ref{eq-01}) works much worse than the AFM model in
Fig.~\ref{fig-06}(a). This suggests that it is not free magnetic
moments (or ferromagnetic inclusions such as MnAs nanocrystals) that
responds to $B$ but rather, an antiferromagnetic system. It could be
that antiferromagnetically coupled pairs of free magnetic moments are
responsible for that but given calculated AMR in Tab.~\ref{tab-04}
it appears likely that we observe bulk response of
an antiferromagnet even if it is probably only a fraction of its
volume (while its substantial part may be strongly pinned by, for
example, structural defects). Another indication that different parts of
the system respond differently to $\vec{B}$ is the non-vanishing saturation
field at $T=300$~K (see the middle panel of Fig.~4). At this temperature,
$B_a$ inferred by SW modelling at saturation nearly
vanishes, yet this should be understood as an effect of averaging two or more
actual sources of magnetic anisotropy rather than its complete suppression.

\section{Conclusion}

Transport properties experimentally investigated in this work are the
magnetoresistance and temperature-dependent resistivity. As for the
latter, we find a reasonable agreement between the large structural
anisotropy (at low temperatures, the out-of-plane resistivity is
almost seven times larger than the in-plane resistivity) and model
calculations which show similar, even if typically somewhat smaller,
anisotropy regardless of the impurity type. This anisotropy is
therefore likely to arise due to layered structure of tetragonal
CuMnAs. We encounter frequent violations of Matthiessen rule:
for varied concentrations of static impurities, for different types of
chemical disorder (at $T=0$) and also for phonons and magnons.
Anisotropic magnetoresistance (AMR) measured is
modest in magnitude and phenomenological modelling indicates the
presence of in-plane uniaxial anisotropy which is not oriented
along any special crystallographic direction. It is at present impossible to
conclude what part of our system responds to the applied magnetic
field but it is unlikely that the single-domain picture applies.

\section*{Acknowledgements}


Discussions with T. Jungwirth and J. Zub\'a\v c are acknowledged as
well as the support from National Grid Infrastructure MetaCentrum
CESNET (No. LM2015042), the Ministry of Education, Youth and Sports
from the Large Infrastructures for Research, Experimental Development
and Innovations project 'IT4Innovations National Supercomputing Center --
LM2015070', NanoEnviCz (No. LM2015073), Materials Growth
and Measurement Laboratory MGML.EU (No. LM2018096, see: {\tt http://mgml.eu})
all provided under the program 'Projects of Large Research, Development, and
Innovations Infrastructures'. I.T., J.K. and D.W. acknowledge financial support
by contract Nr. 18-07172S from GA\v CR, work of J\v Z was supported by
contract No. 19-18623Y of GA\v CR and also by the Institute of Physics
of the Czech Academy of Sciences and the Max Planck Society through
the Max Planck Partner Group programme. P.H. and E.D.N acknowledge
funding by ERDF under the project CZ.02.1.01/0.0/0.0/15\_003/0000485
and also, EU FET Open RIA Grant No. 766566 and Ministry of Education
of the Czech Republic Grant No. LM2018110 and LNSM-LNSpin are acknowledged.

\begin{appendix}

\begin{figure}
\includegraphics[scale=0.5]{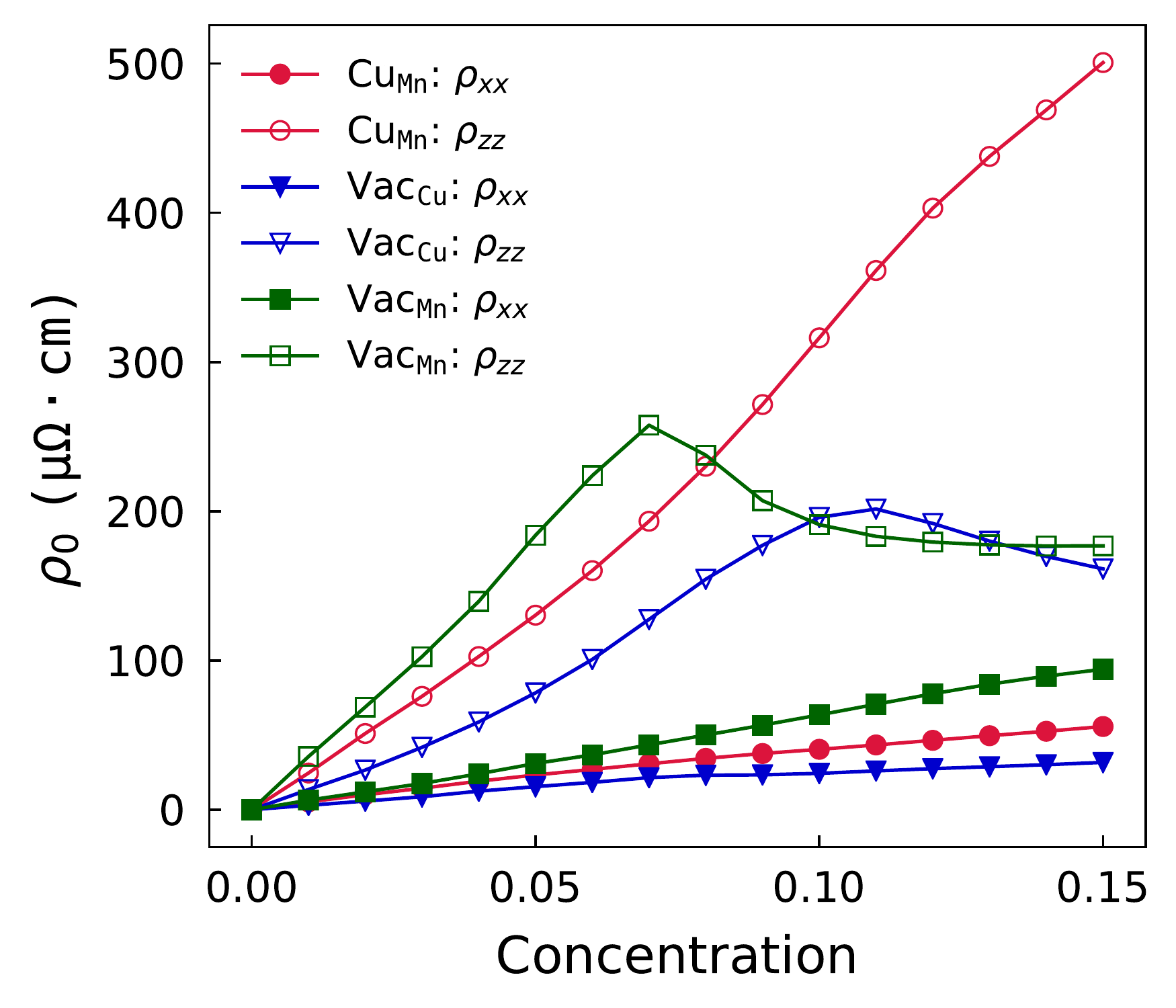}
\caption{Zero-temperature resistivity for three different types of
  impurities as a function of concentration of the respective impurity.}
\label{fig-07}
\end{figure}

\section{Magnetic anisotropies and SW model}

Apart from magnetocrystalline anisotropy energy (MAE), lower than
cubic symmetry systems are affected by dipolar interactions as far as
their easy axes are concerned.\cite{Correa:2018_a}  Using DFT+U
calculations,\cite{Wadley:2015_a} MAE was estimated
at 0.130~meV/f.u. favouring the in-plane directions
and the dipole-dipole interactions, evaluated using Eq.~(A1) of
Ref.~\onlinecite{Correa:2018_a}, further increase the energy penalty
for magnetic moments along $c$-axis by 0.04~meV/f.u.

For $\vec{m}$, $\vec{b}$ lying in-plane, the first term in
Eq.~(\ref{eq-02}) can be rewritten using angles $\psi$, $\varphi$ as
$-B\cos(\psi-\varphi)$ and the magnetic anisotropy using
\begin{equation}\label{eq-04}
2\cos^2(\phi-\phi_0)-1= \cos 2\phi_0 \cos 2\phi + \sin 2\phi_0 \sin 2\phi
\end{equation}
where $\vec{a}=(\cos\phi_0,\sin\phi_0)$ is the easy axis direction.
This allows to immediately identify $\alpha=B_a/B_0$ and $\beta=B/B_0$
in Eq.~\ref{eq-03} in the case of ferromagnets ($B_0$ is a reference
field). For antiferromagnets, 
the derivation of Eq.~\ref{eq-03} with $n=2$ is more involved. First,
the two angles related to $\vec{m}_{1,2}$ are reduced to just one (the
one related to canting, i.e. effectively $\vec{m}_1+\vec{m}_2$ can be
expressed analytically and then re-inserted into Eq.~\ref{eq-01}).
Direction of the N\'eel vector, parametrised by angle $\psi$, remains
as a variable with respect to which the energy should be minimised. 
Eq.~\ref{eq-03} with $n=2$ follows and $\alpha=B_a/B_e$ whereas $\beta=B/B_e$.

Good fits in Fig.~\ref{fig-04} are only possible if we allow for
nonzero $\phi_0$ and, with respect to the [100] crystallographic
direction, we find that $\vec{a}$ is inclined by $\approx 15^\circ$ at
low temperatures. Biaxial anisotropy can be modelled by replacing
$\cos 2\phi$ with $\cos 4\phi$ in Eq.~(\ref{eq-04}) but fits give
a significantly larger $\chi^2$ (about a factor of five) than for
uniaxial anisotropy. The difference in quality of the fits (uniaxial
and biaxial, both with $\phi_0$ as a free parameter) is also clearly visible.

\section{Detailed transport calculations}

Tab.~\ref{tab-01} of the main text summarizes the most important results for
zero-temperature resistivity. However, various approaches (within CPA
based on TB-LMTO) to calculate resistivity can be chosen:
Tab.~\ref{tab-03} gives an overview of
both scalar and fully relativistic approaches and the effect of
Hubbard $U$ and $spd$ vs. $spdf$ basis is also presented. (We note that
data in Tab.~\ref{tab-02} are calculated using the $spdf$ basis.)
The discrepancies among the values in the table should be considered
as an uncertainty of our approach; we note, that a larger basis in the
TB-LMTO does not necessary lead to more precise calculations. 
Resistivities in both Tab.~\ref{tab-01} and \ref{tab-03} are shown for 5\% of
the respective impurity and formation energies are taken from
Ref.~\onlinecite{Maca2019}.
In general, the lowest resistivities are obtained for the scalar
relativistic approach and the values are also larger for the $spdf$
basis; however, there is no strict trend and various impurities behave
differently.

\begin{figure}
\includegraphics[scale=0.5]{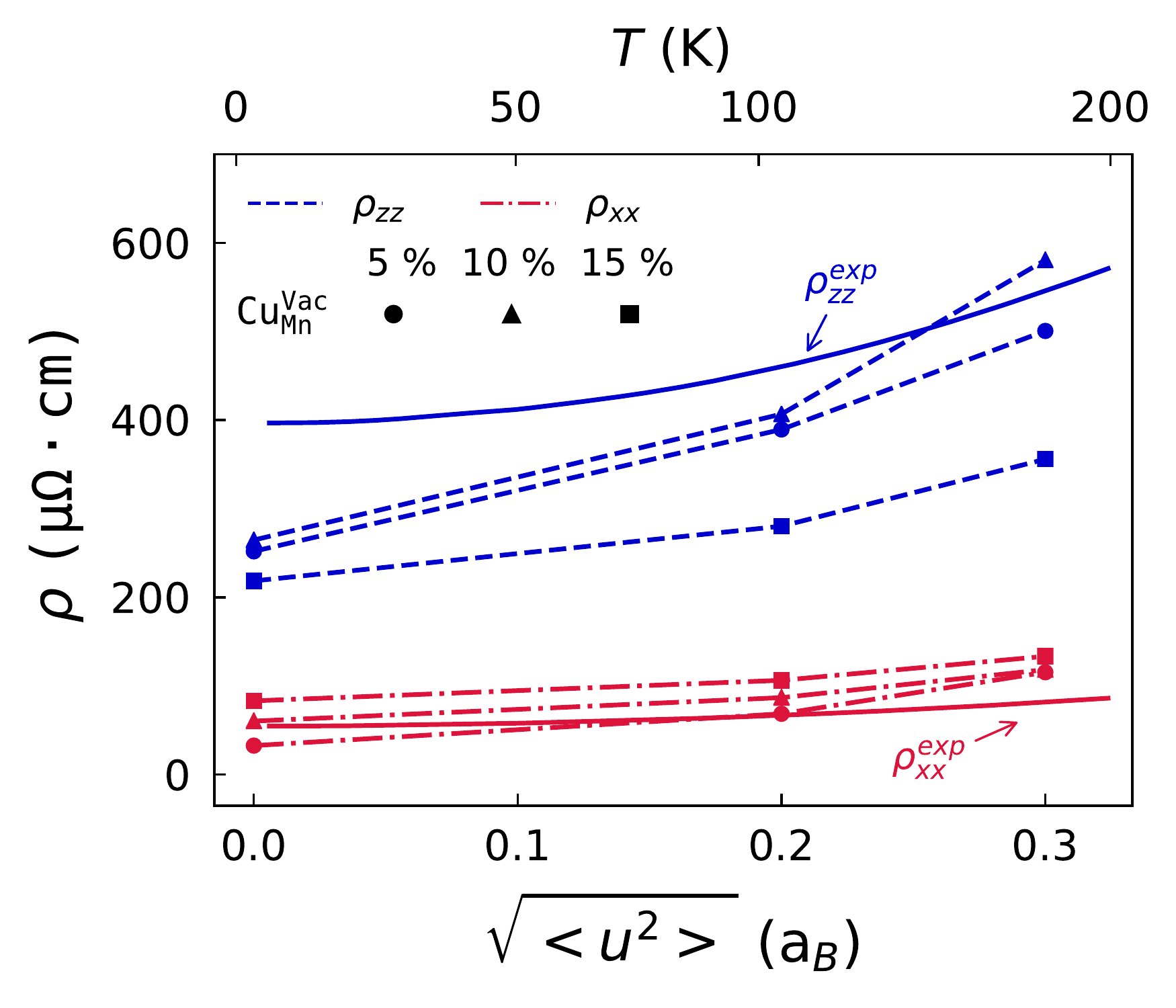}
\caption{More model calculations of $\rho(T)$ compared to experimental
  data (solid lines). Here, the sources of scattering\cite{note3}
  are Cu$^{\mathrm{vac}}_{\mathrm{Mn}}$ and phonons.}
\label{fig-05}
\end{figure}

We proceed with a remark on additivity of scattering rates in the
context of zero-temperature resistivity. Not only that the Matthiessen
rule does not hold for {\em different} sources of scattering; even
with a single type of impurity, doubling its concentration does not
necessarily lead to doubling the resistivity. A clear example of this
is shown in Fig.~\ref{fig-07}. The most striking case is that of
non-monotonic $\rho_{zz}$ with maxima at 7\% and 11\% of Vac$_\mathrm{Mn}$ and
Vac$_\mathrm{Cu}$, respectively. In the context of binary alloys, these
concentrations are relatively low but similar values have been reported for
nonmagnetic Pd-Co\cite{Kudrnovsky:2015_a}
and magnetic Ni-Fe and Ni-Co.\cite{Turek:2012_a}
We note, that since these random alloys are cubic, the anisotropy is
of minor influence there.

Non-monotonic dependence of $\rho_{zz}$ on impurity concentration occurs
also for the more complex model mentioned in Fig.~\ref{fig-01}. As a
consequence, increasing the concentration of\cite{note3}
Cu$^{\mathrm{vac}}_{\mathrm{Mn}}$
does not improve the agreement with experimental data, see Fig.~\ref{fig-05}.
Temperature-dependence of resistance, nevertheless, agrees reasonably well
as far as phonons are concerned and this applies to a larger group of
impurities. Linear function was fitted to $\rho_{xx}(T)$ and $\rho_{zz}(T)$
in the range from 0~K to 180~K and the linear coefficients\cite{note2}
are shown in Tab.~\ref{tab-05}  Negative values of these coefficients are
usually not observed in experiments; nevertheless, measured resistance
may decrease with growing chemical disorder and this is also seen in
our model results of Fig.~\ref{fig-07}. Obtained linear coefficients
for $\rho_{zz}(T)$ (shown in Tab.~\ref{tab-05}) are much more
sensitive to the kind of the impurity than in the case of
$\rho_{xx}(T)$, i.e., the standard deviation of the average value (of
the calculated data) is more than 110 \% for $\rho_{zz}(T)$ while
similar analysis for $\rho_{xx}(T)$ gives standard deviation below
30 \%. Together with formation energies and residual resistivities
(Tab.~\ref{tab-01} and~\ref{tab-03}), the trends may be used to
determine the most probable defects.

\begin{table}
  \begin{tabular}{|l|cc|}\hline
    Defect      &       $\rho_{xx}$ [$\mu\Omega$ cm K$^{-1}$]                   
&       $\rho_{zz}$ [$\mu\Omega$ cm K$^{-1}$]                   \\ \hline\hline
    As$_{\rm{Mn}}$      &       0.32    $\pm$   0.03    &       -0.76   $\pm$   
0.06    \\
    As$_{\rm{Cu}}$      &       0.40    $\pm$   0.02    &       -0.43   $\pm$   0.14    \\
    Mn$_{\rm{As}}$      &       0.37    $\pm$   0.03    &       -0.20   $\pm$   0.03    \\
    Cu$_{\rm{As}}$      &       0.35    $\pm$   0.05    &       0.29    $\pm$   0.11    \\
    Cu$\leftrightarrow$Mn       &       0.45    $\pm$   0.05    &       0.43    $\pm$   0.10    \\
    Vac$_{\rm{As}}$     &       0.34    $\pm$   0.02    &       0.52    $\pm$   0.17    \\
    Mn$_{\rm{Cu}}$      &       0.44    $\pm$   0.05    &       0.68    $\pm$   0.01    \\
    Cu$_{\rm{Mn}}: 10 \%$       &       0.47    $\pm$   0.04    &       1.29    $\pm$   0.07    \\
    Vac$_{\rm{Cu}}$     &       0.48    $\pm$   0.09    &       1.30    $\pm$   0.19    \\
    Vac$_{\rm{Mn}}$     &       0.46    $\pm$   0.07    &       1.33    $\pm$   0.07    \\
    Cu$_{\rm{Mn}}: 5 \%$        &       0.54    $\pm$   0.05    &       1.51    $\pm$   0.09    \\
    Cu$_{\rm{Mn}}: 2 \%$        &       0.62    $\pm$   0.09    &       1.59    $\pm$   0.18    \\
    No impurity &       0.70    $\pm$   0.23    &       1.62    $\pm$   0.41    \\ \hline\hline

    Experiment  &       0.23    $\pm$   0.01    &       1.30    $\pm$   0.02    \\ \hline
  \end{tabular}
  \caption{Linear coefficient from $\rho(T)$ fits up to
    $T=180$~K, uncertainties were obtained from the fit, and
    calculated data ($spdf$, $U=0.00$~Ry) are sorted by the last column.
    The last row shows the same coefficients from the measured values.}
  \label{tab-05}
\end{table}

\begin{table*}[h]
  \begin{tabular}{|c|c|cc|cc|cc|cc|cc|} \hline
        &       Formation       &       \multicolumn{2}{c|}{Scalar rel., \textit{spd}}                  &       \multicolumn{4}{c|}{Fully rel., \textit{spd}}                                                   &       \multicolumn{4}{c|}{Fully rel., \textit{spdf}}                                                  \\
        &       energy \cite{Maca2019}  &       \multicolumn{2}{c|}{$U=0.00\,\text{Ry}$}                        &       \multicolumn{2}{c|}{$U=0.00\,\text{Ry}$}                        &       \multicolumn{2}{c|}{$U=0.10\,\text{Ry}$}                        &       \multicolumn{2}{c|}{$U=0.00\,\text{Ry}$}                        &       \multicolumn{2}{c|}{$U=0.10\,\text{Ry}$}                        \\
Defect  &       [eV]    &       $\rho_{xx}$     &       $\rho_{zz}$     &       $\rho_{xx}$     &       $\rho_{zz}$     &       $\rho_{xx}$     &       $\rho_{zz}$     &       $\rho_{xx}$     &       $\rho_{zz}$     &       $\rho_{xx}$     &       $\rho_{zz}$     \\ \hline
Vac$_\textrm{Mn}$ &       -0.16   &       36      &       155     &       32      &       154     &       19      &       134     &       31      &       184     &       20      &       181     \\
Vac$_\textrm{Cu}$ &       -0.14   &       12      &       44      &       12      &       54      &       9       &       57      &       16      &       79      &       11      &       92      \\
Mn$_\textrm{Cu}$  &       -0.03   &       111     &       171     &       115     &       203     &       132     &       683     &       112     &       263     &       150     &       915     \\
Cu$_\textrm{Mn}$  &       0.34    &       24      &       122     &       22      &       130     &       8       &       40      &       23      &       131     &       8       &       57      \\
Cu$_\textrm{As}$  &       1.15    &       107     &       273     &       109     &       377     &       144     &       989     &       121     &       481     &       163     &       1299    \\
As$_\textrm{Cu}$  &       1.73    &       94      &       219     &       98      &       257     &       112     &       530     &       114     &       359     &       123     &       694     \\
As$_\textrm{Mn}$  &       1.79    &       113     &       262     &       124     &       240     &       133     &       455     &       141     &       476     &       161     &       617     \\
Mn$_\textrm{As}$  &       1.92    &       122     &       151     &       130     &       270     &       155     &       854     &       147     &       423     &       186     &       1784    \\
Vac$_\textrm{As}$ &       2.18    &       174     &       203     &       182     &       246     &       219     &       1054    &       210     &       306     &       284     &       1556    \\
Cu$\leftrightarrow$Mn   &       -       &       124     &       267     &       123     &       304     &       127     &       629     &       120     &       393     &       142     &       882     \\ \hline
  \end{tabular}  
  \caption{
    Detailed microscopic calculations of
    resistivities in $\mu\Omega\cdot$cm for 5\%
    of the respective impurity.}
\label{tab-03}
\end{table*}

\begin{figure}
\includegraphics[scale=0.5]{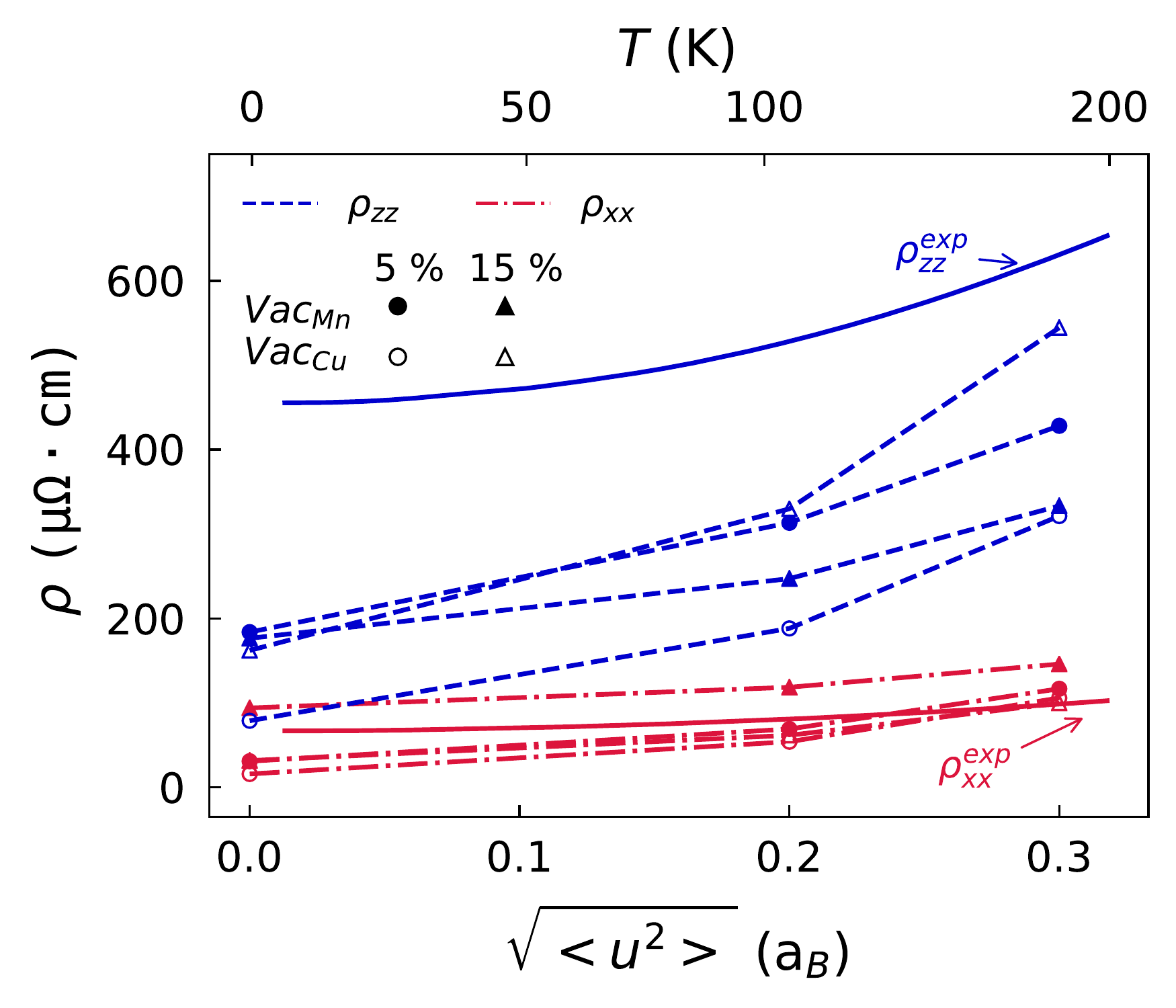}
\caption{Model calculations analogous to Fig.~\ref{fig-05} with 
    vacancies on Cu and Mn sites instead of Cu$^{\mathrm{vac}}_{\mathrm{Mn}}$.}
\label{fig-09}
\end{figure}

To give another example of phononic effects, we show
temperature-dependent resistivity for vac$_{\mathrm{Cu}}$ and
vac$_{\mathrm{Mn}}$ in Fig.~\ref{fig-09}. Note that the linear
coefficients of $\rho_{zz}(T)$ in Tab.~\ref{tab-05} are in a very good
agreement with experimental values for these impurities. Combining
Fig.~\ref{fig-09} with Fig.~\ref{fig-08} leads to different
resistivities than what is shown in Fig.~\ref{fig-05} thus
demonstrating the failure of the Matthiessen rule once again.

\begin{figure}
\includegraphics[scale=0.5]{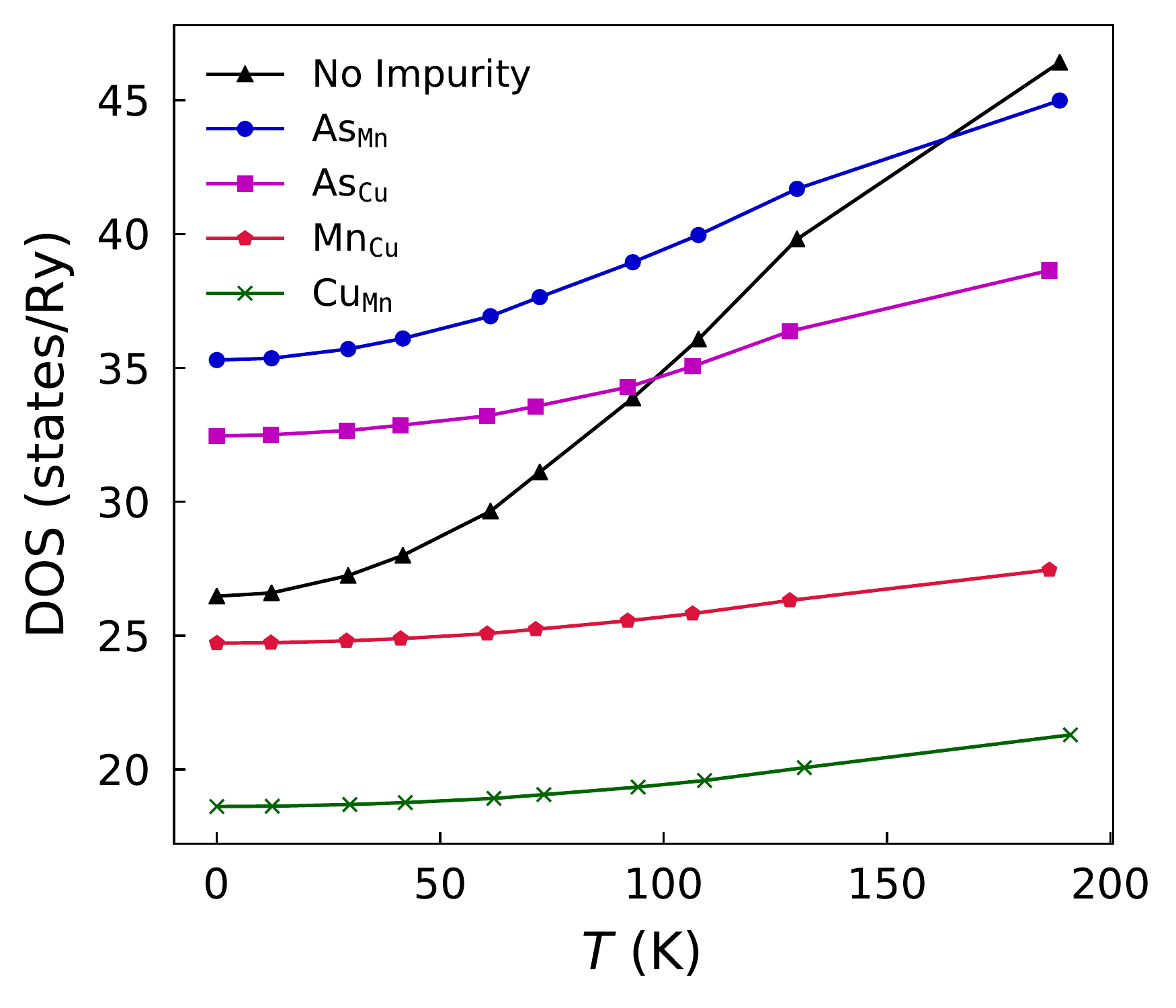} 
\caption{Total DOS at $E_F$ with only phononic contribution to temperature.
CuMnAs with no impurities is shown by black line with crosses and 5~\% of As$_{\rm{Mn}}$, As$_{\rm{Cu}}$, Mn$_{\rm{Cu}}$, and Cu$_{\rm{Mn}}$ is depicted by gray circles, blue triangles, green squares, and red diamonds, respectively.}
\label{fig-11}
\end{figure}

We conclude this appendix by several comments on the correlation of
resistivity to the density of states (DOS) at the Fermi level $E_F$.
The saturation of $\rho_{xx}(T)$ and the decrease of $\rho_{zz}(T)$
(with increasing temperature) caused by magnons was attributed
in Ref.~\onlinecite{DW2019-JMMM} to a high increase of DOS at the Fermi level. 
Here we observe decreasing $\rho_{zz}(T)$ due to phonons for some
impurities but $\rho_{xx}(T)$ having reasonable metallic-like
increase. It is shown in Tab.\ \ref{tab-05} (negative slopes) and in
Fig.~\ref{fig-08} and we also tried to address it on the level of the DOS.
(The energy dependent DOS were calculated, but they are not shown here
for brevity; they are presented in Refs.~\onlinecite{DW2019-JMMM,Maca2019}).
For clean stoichiometric CuMnAs, the DOS is strongly increasing above $E_F$,
i.e., there are about twenty states per Ry at $E_F$, while four times
more for $E>E_F + 0.2$~eV. Under the presence of phonons, 
this region of high DOS is smeared (more precisely: large self-energy
leads to a large broadening of the spectral function) and for the
stoichiometric CuMnAs, situation at $E_F$ is appreciably modified
for $T\gtrsim 100$~K (see the black line with crosses in Fig.~\ref{fig-11}).
The drop of $\rho_{zz}(T)$ in Fig.~\ref{fig-08} begins around 200~K
which can be expected given the fact that the increase of DOS with
temperature is initially compensated by an increase of self-energy.
We note that no similar decrease with temperature is observed
for $\rho_{xx}(T)$; this could be caused by the layered structure of CuMnAs,
but directionally resolved study of the states, e.g., in the terms of
the Bloch spectral function similarly to previously investigated
NiMnSb\cite{DW2019-PRB}, is beyond the scope of this paper.
Although we attribute the phonon-induced decrease of resistivity to
the DOS specific for CuMnAs, it could occur also for other metals
having similar DOS.

\end{appendix}


\begin{thebibliography}{99}
\bibitem{note3} Cu$^{\mathrm{vac}}_{\mathrm{Mn}}$ denotes a 'combined' impurity
  where a Cu atom is removed from its original position and put at Mn site,
i.e. effectively a combination of Cu$_{\mathrm{Mn}}$ and vac$_{\mathrm{Cu}}$
single-atomic impurities.
\bibitem{note1} To be precise, we measured $\rho_{xx}$ for
  different directions of $\vec{B}$ rather than $\rho_{xx}$ different
  from $\rho_{yy}$ for the same $\vec{B}$. Given the tetragonal
  symmetry, however, the two cases are directly related.
\end{thebibliography}

\end{document}